# Study of Growth Properties of InAs Islands on Nucleation Sites Defined by Focused Ion Beam

R-Ribeiro Andrade*[1,6], D. R. Miquita[1], T. L. Vasconcelos[2], R. Kawabata[4,6], A. Malachias[5], M. P. Pire[3,6], P. L. Souza[4,6] and W. N. Rodrigues[5,6]

[1] Centro de Microscopia, UFMG, Belo Horizonte, MG, Brazil
[2] Divisão de Metrologia de Materiais, INMETRO, Duque de Caxias, Brazil
[3] Instituto de Física, UFRJ, Rio de Janeiro, Brazil
[4] LabSem/CETUC, PUC-Rio, Rio de Janeiro, Brazil
[5] Departamento de Física, ICEx, UFMG, Belo Horizonte, MG, Brazil
[6] DISSE – Instituto Nacional de Ciência e Tecnologia de Nanodispositivos Semicondutores, CNPq/MCT, Brazil

*corresponding author: rodriban@fisica.ufmg.br

**ABSTRACT**

This work describes morphological and crystalline properties of the InAs islands grown on templates created by focused ion beam (FIB) on indium phosphide (InP) substrates. Regular arrangements of shallow holes are created on the InP (001) surfaces, acting as preferential nucleation sites for InAs islands grown by Metal-Organic Vapor Phase Epitaxy. Ion doses ranging from $10^{15}$ to $10^{16}$ Ga$^+$/cm$^2$ were used and islands were grown for two sub-monolayer coverages. We observe the formation of clusters in the inner surfaces of the FIB produced cavities and show that for low doses templates the nanostructures are mainly coherent while templates created with large ion doses lead to the growth of incoherent islands with larger island density. The modified island growth is described by a simple model based on the surface potential and the net adatom flow to the cavities. We observe that obtained morphologies result from a competition between coarsening and coalescence mechanisms.

## 1. INTRODUCTION

Quantum dots (QDs) are atom-like carrier-confining nanostructures that can be used in high efficiency LEDs, lasers and photodetectors, among other applications. The role of QDs in the efficiency gain of such devices comes from the three-dimensional confinement of the charge carriers resulting in a discrete distribution of their energy levels. Tuning shape, size, density, size distribution and composition of the QDs allow the design of devices for very specific applications [1,2]. However, to achieve this possibility a fine control of deposition/growth variables is mandatory. Despite of the capabilities of the usual growth techniques, which provide a good control of the composition, the adsorption of precursors on multifaceted or rough surfaces may occur as a sum of distinct surface-mediated mechanisms, turning the process uncontrollable. Growing in such scenario may lead to uncontrolled properties of the QDs. Therefore it is of paramount importance to have a fine control of the nucleation sites and their morphologies, allowing the manipulation of morphological, structural and chemical properties of QDs which directly impact on their electronic states [1,2]. To control growth processes under such conditions mentioned above a combination of techniques to induce QDs localized growth is required. Among the known patterning techniques one can list lithography using ultraviolet light [3,4], X-rays [4] or electron beam [5], as well as indentation and local chemical changes induced via scanning probe microscopy [6]. Alternatively, the Focused Ion Beam technique (FIB) is a very promising tool to induce site-controlled nucleation of QDs with an accuracy of a few tens of nm [7,8]. With FIB it is possible to implant Ga (or other) ions on a variety of substrates in specific locations to create favorable nucleation sites for the QDs growth [2,7] or generate distinct surface profiles to grow QDs [8-10]. Since the feasibility of using FIB to produce site-controlled InAs QDs on GaAs substrates was demonstrated efforts have been driven towards the production of devices based on this technology [2,7-9]. Despite the advances on this method, the dependencies of $Ga^+$ dose on shape, size distribution, defects, chemical composition and interdiffusion of the atomic speciments in nanostructures remains poorly understood. In the present investigation InAs islands were grown by Metal-Organic Vapor Phase Epitaxy (MOVPE) on InP (001) surfaces previously patterned by $Ga^+$ focused ion beams, producing a regular arrangement of shallow cavities on the substrate surface. Subsequently InAs islands were found to grow on the cavity sites, and their properties were analyzed using scanning electron microscopy (SEM), transmission electron microscopy (TEM), energy dispersive spectroscopy (EDS), grazing incidence X-ray (GID) and Raman Spectroscopy techniques. A simplified model based on the island density dependence on the Gaussian depth profile of the cavity has been developed, allowing the understanding of the mechanism behind the observed modified island growth.

## 2. EXPERIMENTAL METHODS

A dual-beam Quanta® 3D FEI was used to create an array of shallow cavities on the InP surface prior to the deposition of InAs islands by MOVPE. Initially the InP substrates were cleaned in ultrasonic bath with trichlorethylene (TCE) (3 minutes), acetone (3 minutes) and isopropyl alcohol (3 minutes). The substrates were then dried under ultrapure nitrogen flow and introduced into the FIB chamber. In order to generate the templates, the ion beam was held stationary at a single point of the InP surface for a given time interval. After which the ion beam was regularly shifted by 500 nm along a square lattice to a new point and remained there for the same time interval. This process was repeated 200 x 200 times and a square pattern of shallow cavities was created. The whole pattern was regularly repeated along the InP sample surface 21 x 21 times. The $Ga^+$ ion beam current was set to 60 pA and the ion accelerating voltage to 30 kV. Beam incidence time intervals equal to 0.45 ms, 1.00 ms and 2.30 ms were used, yielding $3.7 \times 10^{15}$, $5.6 \times 10^{15}$ and $1.3 \times 10^{16}$ $Ga^+/cm^2$ doses, respectively. Current and time values were chosen in order to produce cavities with diameters sizes similar to average InAs islands sizes grown on a flat InP surface [10]. The volume of each cavity determined by AFM was $2.5 \times 10^{-4}$ $\mu m^3$, $4.3 \times 10^{-4}$ $\mu m^3$ e $7.1 \times 10^{-4}$ $\mu m^3$ for the doses mentioned above, respectively.

Following the production of the templates on the InP(001) surfaces InAs islands were grown using MOVPE. Before introducing the samples into the MOVPE Aixtron-AIX200 reactor they were cleaned in isopropyl alcohol at 50 °C for 5 min and by oxygen plasma during 30 s. The surface oxide was removed by immersing the samples in 1 % sulphuric acid for 6 s. All substrates were then washed in deionized water for 1 min and finally rinsed in isopropyl alcohol at 50 °C for 1 min. They were then dried in nitrogen flow and immediately introduced into an growth chamber. Once inside the reactor they were heated to 520 °C under a phosphine ($PH_3$) rich atmosphere with a 100 sccm flow (kept for 2 min). The phosphine flow was then replaced by a 10 sccm arsine flow ($AsH_3$) for 1 s and a 16.5 sccm flow of trimethyl-indium (TMIn) with a V/III ratio of 92. The exposure time of the InP surface to TMIn was 3 s or 4 s (depending on the sample series) and is hereafter referred as the growth time. Under these conditions, 0.3 ML and 0.4 ML of InAs were deposited, respectively, with no evidence of wetting layer formation on the surface [6]. For these growth times, islands are large enough to be observed by SEM. After the deposition time the TMIn source was closed and, 12 s later, the sample was cooled down from 520 ºC to 350 °C under $AsH_3$ flow. Below 350 °C the $AsH_3$ flow was suppressed and the samples cooled down to room temperature.

Six different samples were prepared following the procedure described above with the doses previously specified and growth times of 3 s and 4 s, resulting in nominal coverages of 0.6 and 0.8 InAs monolayers, respectively. The growth conditions for all samples are listed in Table I, where the acroyms LD, MD and HD refer to the low, medium and high gallium ion dose, respectively. The additional codes T3 and T4 refer to the 3 and 4 s growth time, respectively.

Chemical analysis was performed by Electron Dispersive Spectroscopy using a FEI Titan 80-300 kV Cs-corrected transmission electron microscope operating at 300 kV accelerating potential and equipped with EDS-EDAX detector (132 eV energy resolution). In this method a coherent focused probe is raster scanned across the specimen and the resultant X-ray emission spectrum recorded at each probe position. Atomic Force Microscopy (AFM) measurements were carried out using a NT-MDT Solver Pro microscope, operating under intermittent contact using PointProbe® Plus Non-contact High Frequency (PPP-NCHR) tips with typical 7 nm tip radius. Finally, grazing incidence X-ray diffraction (GID) experiments were performed in the XRD2 beamline of the Brazilian Synchrotron Light Laboratory (LNLS) to retrieve crystalline strain status of the grown InAs islands.

Table I: Growth conditions of the samples

| Sample Code | $Ga^+$ Dose ($\Phi$) ($x10^{15}$ $Ga^+/cm^2$) | Growth time (s) |
|---|---|---|
| LDT3 | 3.7 | 3 |
| MDT3 | 5.6 | 3 |
| HDT3 | 13.2 | 3 |
| LDT4 | 3.7 | 4 |
| MDT4 | 5.6 | 4 |
| HDT4 | 13.2 | 4 |

## 3. RESULTS

All samples were initially studied by AFM. Fig. 1 (a) illustrates the process of the array of holes fabrication by FIB. Fig. 1 (c), (e) and (g) show AFM topographic images of the InP surface containing the patterns manufactured by FIB with the doses 3.7 x $10^{15}$ $Ga^+/cm^2$, 5.6 x $10^{15}$ $Ga^+/cm^2$, e 1.3 x $10^{16}$ $Ga^+/cm^2$, respectively. From an analysis of these images the average roughness of the flat areas was found to be 1.5 nm. Linear profiles passing through the middle of a column with holes are shown in Fig. 1 (b), (d) and (f), where the height axis is scaled for optimal visualization of hole profiles. For each line scan a Gaussian fit is shown, indicating that the shape of these areas can be well described by such function [11,12]. The depth and diameter of the holes were found to vary with the total ion dose. Average depths of 5.3 nm, 7.5 nm and 10.5 nm are found for the lowest, middle and highest dose used, respectively. Diameters of 173 nm, 192 nm and 207 nm are retrieved from the lowest to highest dose, showing that holes are shallow, with diameter/height aspect ratio ranging from 20 to 30.

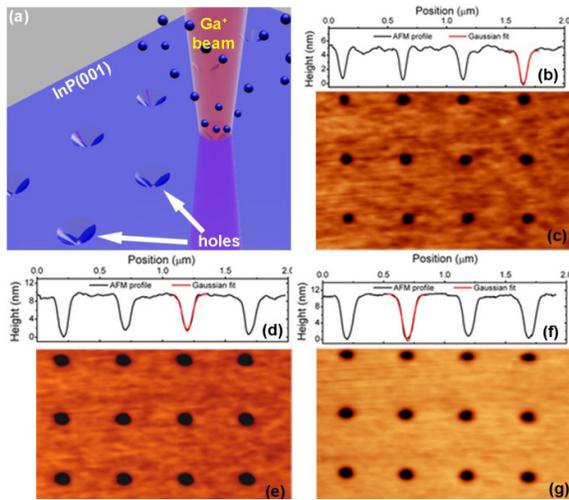

Fig. 1: (a) Sketch of the fabrication process of holes by FIB. AFM topographic images [panels (c), (e) and (g)] show the surface of InP substrate with arrays of holes. From each AFM image linear height profiles were extracted, as shown on panels (b), (d) and (f). The hole profiles is approximately Gaussian, as shown by the red line fits in these panels.

Fig. 2 shows SEM images of different samples, evidencing that InAs islands grow on the FIB-generated templates. Most of the islands are found on the lateral walls of the cavities, while islands covering the entire cavity volume or a large fraction of it were found mostly on the low-dose templates. Islands grown in high-dose templates were found to be smaller in size, forming clusters within the cavities (Fig. 2 (e) and (k)).

The fact that no appreciable difference is retrieved on the roughness by analysis of AFM images inside the holes (they all present smooth surfaces) and their lateral slope is similar indicates that this is a size-dependent effect. High dose templates present holes with larger diameter and in a scenario where nucleation starts at hole edges the amount of deposited material may not be sufficient to fill up the center of such cavities. The lateral island size along the $[\bar{1}10]$ direction and the density of islands were measured by SEM on six different regions of the samples, yielding a total statistical sampling of 1500 islands for each FIB dose. The results of this analysis are shown on the histograms of Fig. 2 Poisson curves were fitted to the histogram data, serving as guide to the eyes solely. The histograms reveal a bi-modal lateral size distribution for the islands grown on a low-dose pattern. In Fig. 2 (b) and (h), the peak density of InAs islands is found to be at 25 nm diameter The second group of islands, with larger lateral. size, is found to be around 125 nm lateral size. Small islands grown on medium doses templates are majoritary and only an extremely reduced amount of the islands on the histograms of Fig 2 (d), (f), (j) and (l) correspond to larger islands (diameter of 100 nm or more). The islands grown on high dose templates are small in size and form clusters within the cavities [Fig. 2 (e), (f), (k) and (l)]. The overall scenario revelead by Fig. 2 points to the fact that the number of islands with lateral size larger than 100 nm decreases significantly increasing the ion dose. Finally, SEM results show no significant difference in morphology between 3 s and 4 s samples indicating a weak dependence of island morphology on the total coverage in the deposition range used in this work.

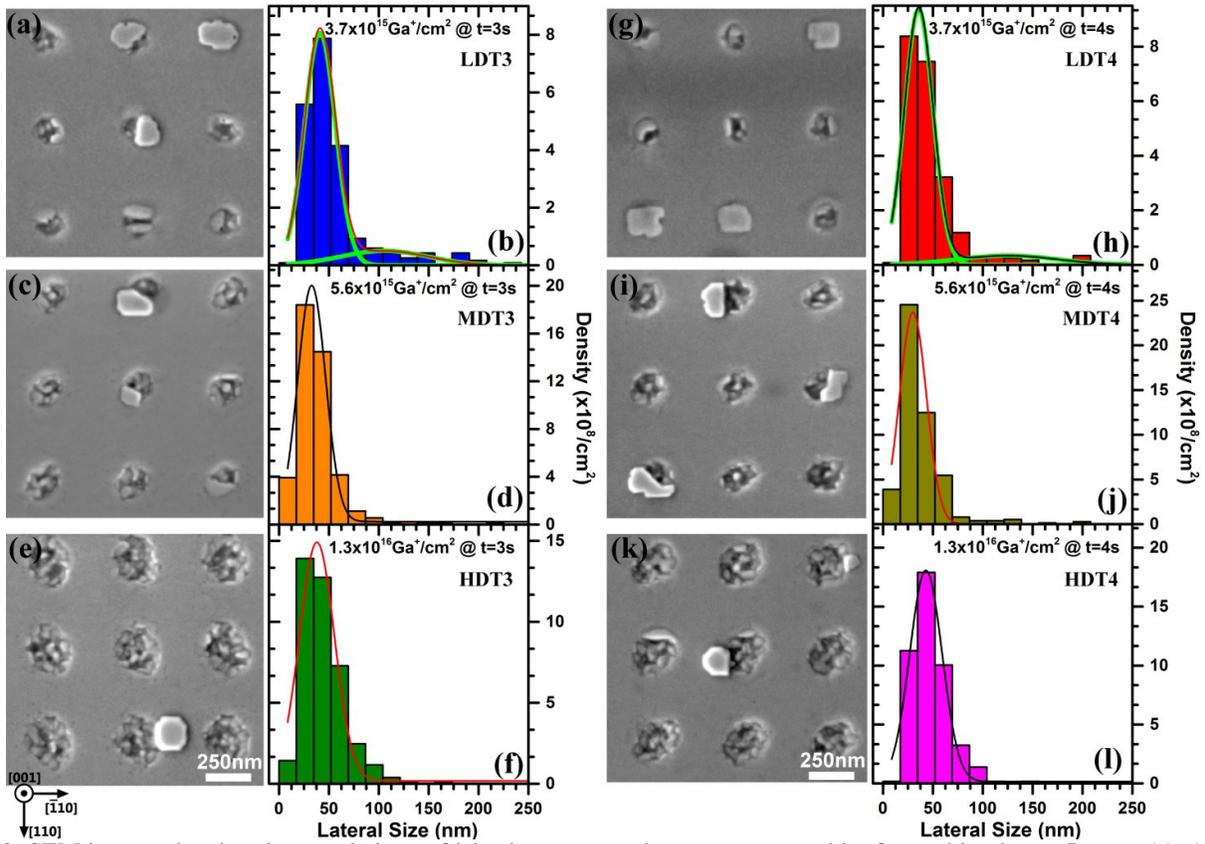

Fig. 2: SEM images showing the morphology of islands grown on the patterns created by focused ion beam. Images (a), (c) and (e) stand for 3s-deposition samples listed on Table I, respectively. Images (g), (i) and (k) correspond to 4s growth time. Lateral size histograms depicting the island diameter distribution on each sample are shown at the right side of SEM images. The $[\bar{1}10]$ direction is oriented along the horizontal axis in all images.

The overall scenario revelead by Fig. 2 points to the fact that the number of islands with lateral size larger than 100 nm decreases significantly increasing the ion dose. Finally, SEM results show no significant difference in morphology between 3 s and 4 s samples indicating a weak dependence of island morphology on the total coverage in the deposition range used in this work. In order to investigate more precisely the morphology of InAs islands and identify their crystalline structure and chemical composition TEM measurements were carried out. All samples studied in cross section mode were prepared by FIB using a technique known as "lift-out" [14]. This method allows us to prepare electron beam transparent lamellas from specific regions of the samples, where the islands were located, and transfer them from the FIB system to the TEM. Fig. 3 (a) shows a TEM image of an island of the LDT4 sample aligned in the [110] zone axis. The island shown has a shape similar to an inverted pyramid and fills the entire volume of the cavity created by the ion beam. EDS maps in scanning transmission microscopy mode (STEM) were carried out to confirm the chemical composition of the island and an elemental mapping was obtained to evaluate the In, P and As distribution. In Fig. 3 (b, c, d), 40 x 20 pixels were acquired with a dwell time of 2 s corresponding to an area of 250 nm x 125 nm. The results confirm that the islands are essentially InAs although the Ga and P. presence cannot be completely discarded if their atomic concentration is below the detection limit of the technique (ranging from 3 to 5%). While the P presence would be an indication of interdiffusion taking place during growth in MOVPE, a small Ga concentration may exist, resulting from the implantation process during the creation of the cavities on the FIB [15].

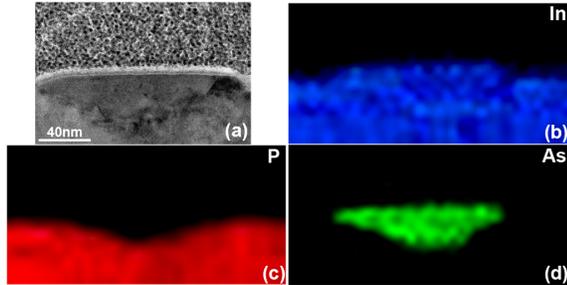

Fig.3: (a) TEM image of an island grown inside a cavity carved out in InP using 3.7 x $10^{15}$ Ga$^+$/cm$^2$ dose (LDT4). The island is imaged laterally along the [110] zone axis. The composition of this island was identified by mapping the elements In, P and As, as shown in images (b), (c) and (d), respectively.

High-Resolution TEM (HRTEM) images of these islands were made using a FEI Tecnai G 20 microscope operating at 200 kV to investigate their lattice integrity as well as defect formation due to the existence of grain boundaries after coalescence of neighboring islands. Fig. 4 shows the results for the LDT4 and HDT3 samples. The panels (a) and (b) show SEM images of the islands for which a detailed TEM analysis is carried out. The small (c), (d), (e) and (f) indicate the panel which shows the correspondent TEM image. Panels 4 (c) and 4 (e) show detailed HRTEM views of islands grown in small clusters on the inner walls of the cavities. Fourier Transforms (FFTs) were also carried out for these images (see small insets), allowing a direct measurement of the crystalline structure. Some island images exhibit Moiré fringes, which are a signature of the interference of structures with different crystallographic orientations with respect to each other and with respect to the substrate. In addition to the Moiré fringes, the HRTEM images also evidence the existence of crystalline defects in these islands. Islands growing oriented under strict epitaxial conditions with the substrate (with lattice parameter matching at the interface) registry were also observed, as shown in Fig. 4 (d) and (f). A reduced amount of such coherent island type was observed with respect to misoriented/relaxed islands. The island shown in Fig. 4 (d) is an example of single-crystalline island orientation, as observed on the FFT inset [Fig. 4 (h)], and possibly corresponds to an epitaxial island. The island in the Fig. 4 (f) exhibit crystalline defects which allow some parts to grow misoriented with respect to the substrate. In consequence, additional spots can be observed in the FFT image (inset) of Fig. 4 (j). The crystallographic characteristics observed in the islands shown in Fig. 4 (d) and (f) were also observed in islands morphologically similar to those, shown in Fig. 4 (c) and (e). These morphological and crystallographic characteristics suggest that the islands grown on FIB templates accommodate their lattice mismatch strain energy by nucleating in the vicinity of defects created by the ion beam in the inner surface of the cavity. On the other hand, when these islands grow and coalesce, the total energy due to the difference in lattice parameter of the materials increases considerably. While small islands accommodate the strain energy easily and can grow misoriented, large islands that grow following the substrate orientation relax due to the formation of defects inside the InAs lattice (Fig. 4 (d) and (f)). All large islands analyzed in this work are morphologically similar to those shown in Fig. 4 characteristics, occurring predominantly on the low-dose templates. All small islands observed by TEM in clusters on the inner walls of cavities [as shown in the Fig. 4 (c) and (e)] are misoriented (d) and (f), showing the same crystallographic with respect to the substrate, in agreement with the InP surface defect scenario previously described.

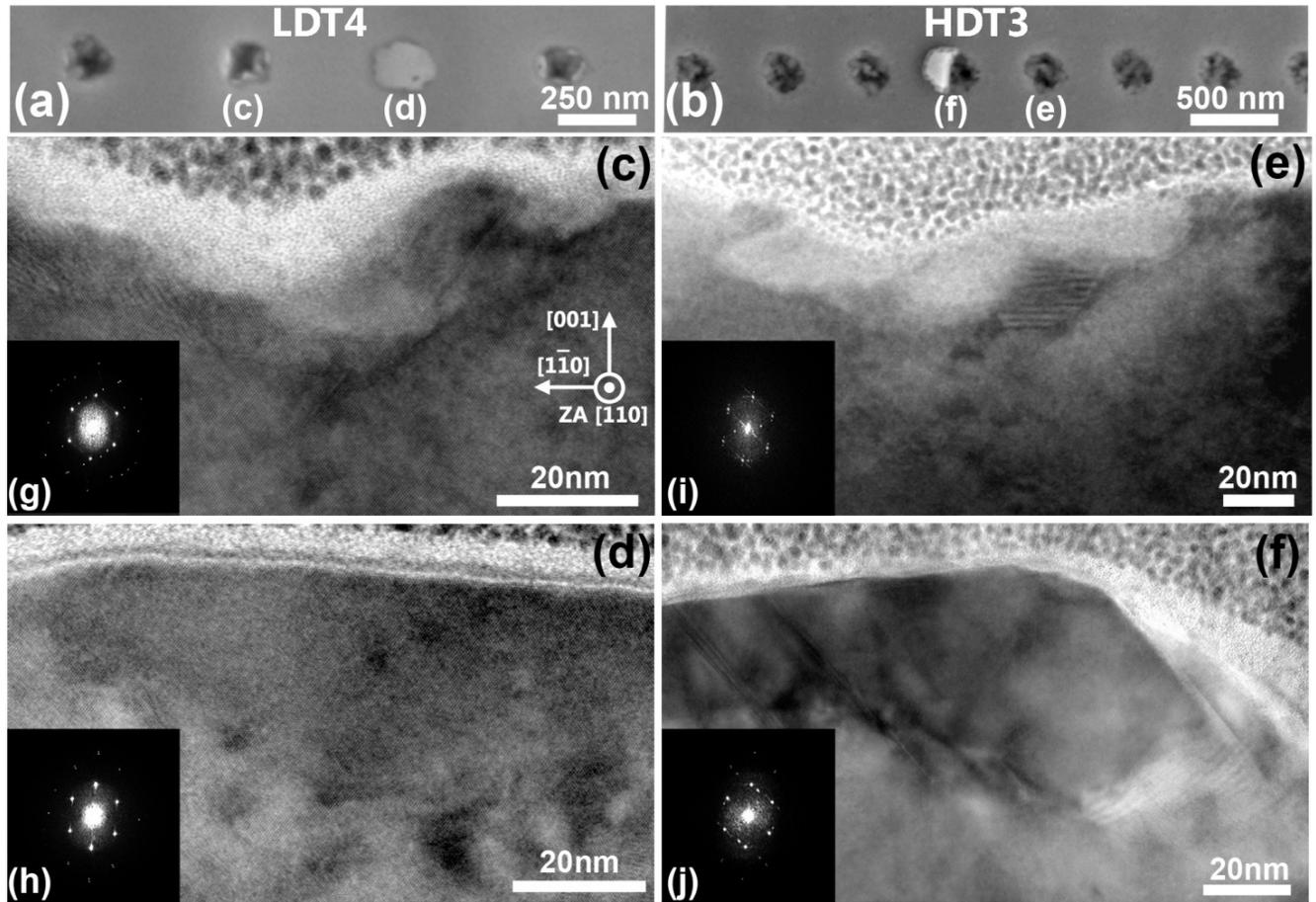

Fig. 4: TEM images from samples LDT4 e HDT3. The islands are aligned at [110] zone axis. Fig. 4(a) and (d) show SEM images of the islands that were chosen for TEM analysis. FFT results corresponding to InAs regions of each TEM image are presented as insets.

Additionally, since InAs islands grown in low-dose templates have lower density and are larger in size (see histograms of Fig. 2) one can conclude that the overall InAs coherent volume is larger for templates prepared with low ion doses.

The Geometric Phase Analysis (GPA) technique developed by Hÿtch et al. [16,17] was used for mapping strain fields inside our islands. The GPA analysis was carried out using the STEM CELL package [18] and measuring the local displacement with respect to a reference region. In order to characterize the local lattice deformation, the **g**=[002] and **g**= $[1\bar{1}\bar{1}]$ were selected. The main limitation of GPA technique is related to the delocalization effects caused (mainly) by the spherical aberration of the objective lens. Maps of in-plane strain fields ($\varepsilon_{xx}$ – horizontal and $\varepsilon_{yy}$ – vertical) were directly converted into lattice parameter maps. Fig. 5 (a) and (d) show HRTEM images extracted from Fig. 4 (d) and (f), respectively. The corresponding strain maps are shown in Fig. 5 (b) - (f). Despite of GPA limitations, the error in local lattice parameter is about ±0.03Å estimated from the local oscillations in the strains curves obtained from vertical scans in the Fig. 5 (b) - (f). The strain maps of Fig. 5 (a) and (e) reveal inhomogeneities on the lattice parameter profile along InAs islands. Abrupt changes observed in Fig. 5 (b) - (f) point to the presence of defects However, large areas on the regions depicted exhibit lattice parameters around (6.02 ± 0.03) Å, which indicate that large islands are able to relax even when they do not nucleate on top of FIB-generated defects. X-ray diffraction measurements under grazing-incidence (GID) conditions were performed in the vicinity of the InP (220) reflection for all samples previously described. Fig.s 6 (a) and (c) show radial (longitudinal) scans, providing a direct evidence of the existence of distinct lattice parameters parallel to the surface, associated to the diffracting objects. In these graphs, the horizontal axis was directly converted into lattice parameter to allow a visual comparison of the position of diffraction peaks and the volume of the diffracting objects, as well as to infer their epitaxial relation with respect to the substrate.

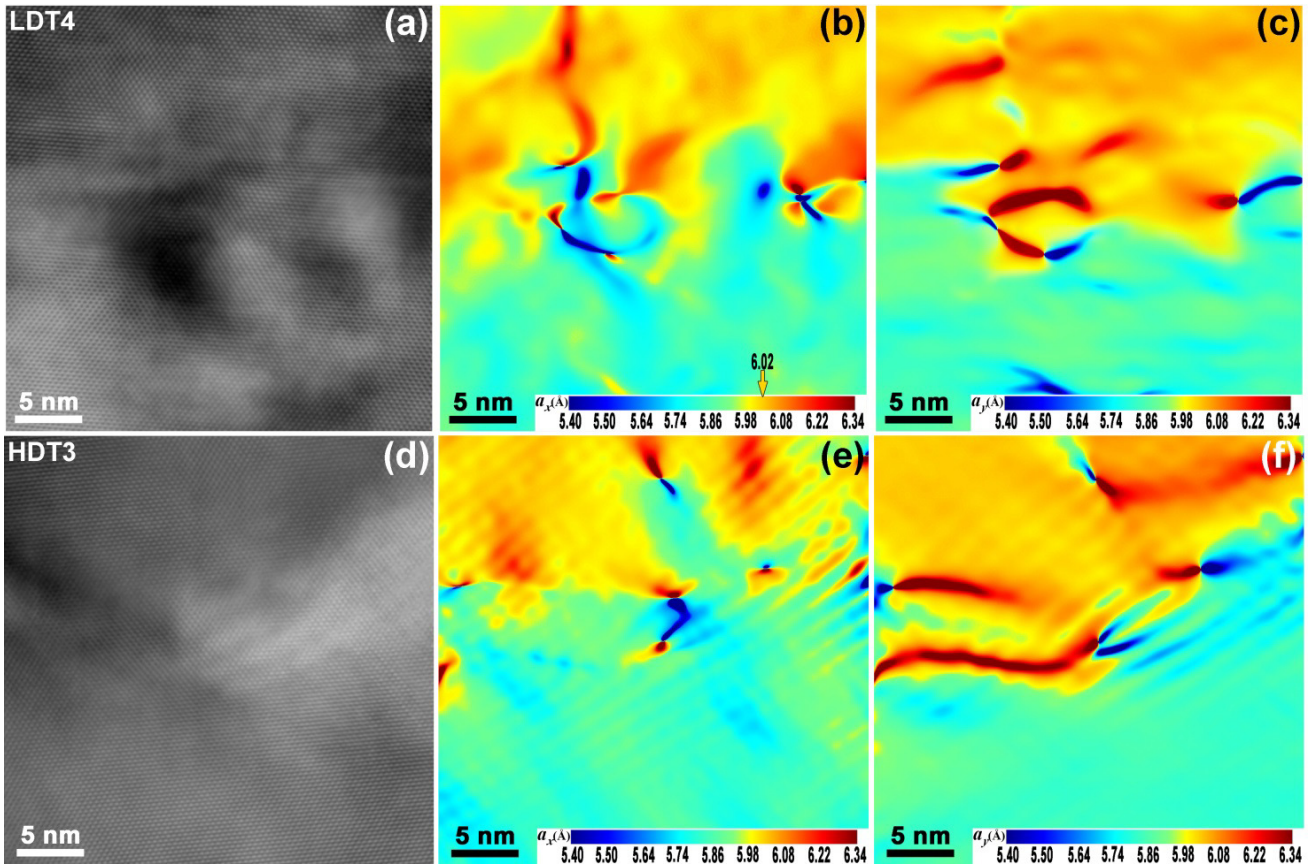

Fig. 5: High resolution TEM images used to produce the lattice parameter maps for the LDT4 (a) and HDT3 (d) samples. Fig. (b) and (c) correspond to the lattice parameter maps along the *x*-direction and *y*-direction, respectively, for the LDT4 sample. The arrow on the scale bar in (b) indicates the average value for the lattice parameter on the InAs island. Fig. (e) and (f) correspond to the *x*-direction and *y*-direction maps, respectively, for the HDT3 sample.

The broad structures around 6.03Å in Fig. 6 (a) and (c) are related to fully relaxed InAs islands. This value is similar to the one obtained from the GPA analysis shown in Fig. 5. A higher intensity near the InAs reciprocal space position is observed in the low-dose sample when compared to the high-dose one [Fig. 6 (a) and (c)]. For samples grown on low-dose templates, with both growth times, the XRD measurements exhibit an increase in the intensity of the InAs peak. This result suggests that the volume of oriented InAs islands increases for low-dose templates, even if the samples produced in these configurations have a reduced island density as shown in Fig. 2. Such result points to a relative improvement of the crystalline orientation of InAs islands under these growth conditions, in agreement with the TEM observations.

In order to study the island lateral size dependence with local lattice parameter and investigate the possibility of existing strain gradients angular (transversal) scans were also performed for reciprocal space conditions that correspond to lattice parameters ranging from InP to InAs. Pseudo-voigt functions were used to adjust the intensity and width ($\Delta q_a$) of all curves [19-22] the local lateral size ($L = 2\pi/(\Delta q_a)$) extracted is shown in the Fig. 6 (b) and (d). One observes that the lateral size increases with growth time and is abruptly for values close to the InP due to the proximity for values close of the substrate peak (related to an infinitely large crystal). For the lattice parameter values close to InAs the crystallite larger for low FIB doses.

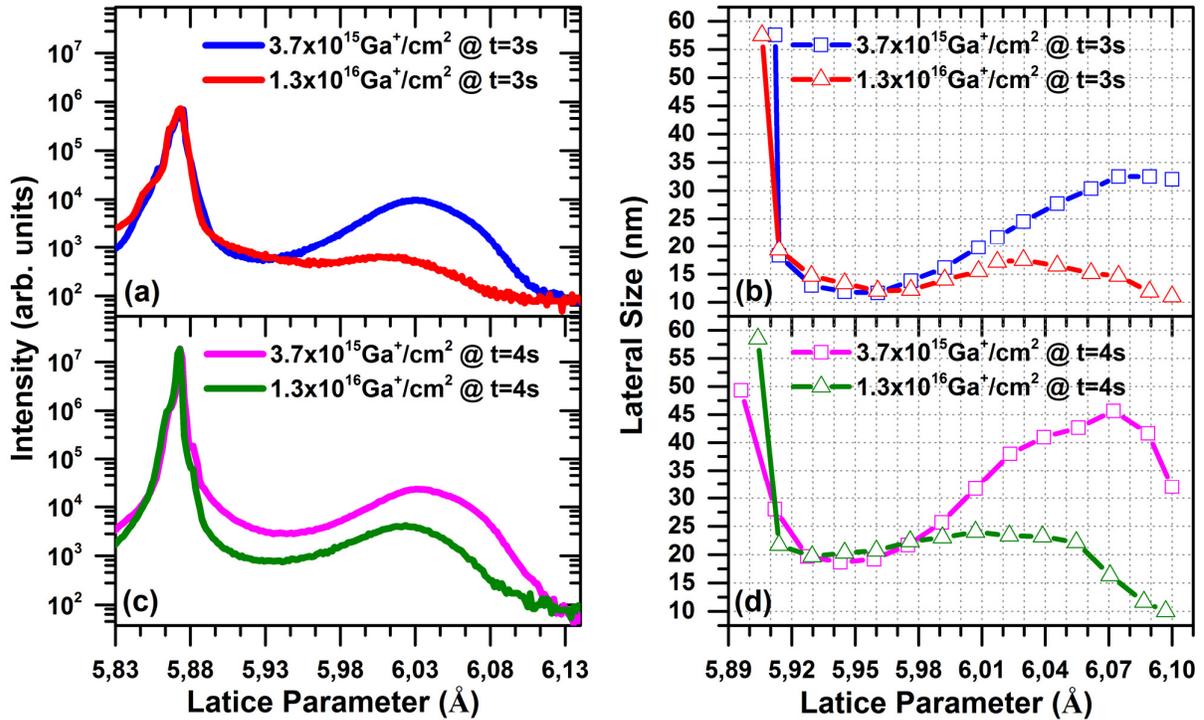

Fig. 6: Grazing-incidence diffraction results for low and high doses samples in the vicinity of the InP (220) reciprocal space position. Panels (a) and (c) show radial scans along the [110] direction. The sharp peak on the left corresponds to the InP substrate the broad peak on the right side is ascribed to InAs islands. In (b) and (d) the results for the local lateral size of the diffracting objects obtained from the widths of angular scans are presented.

Lateral sizes ranging from 10 to 25 nm are retrieved for islands grown on high dose templates [Fig. 6 (b) and (d)] for InAs as well as for intermediate lattice parameters values, indicating that regions inside the islands are subjected to a partial epitaxy (nor fully strained or fully relaxed) [21]. For the low dose templates the lateral size increases considerably near lattice parameter values close to InAs, showing that larger islands with epitaxial registry with the substrate are mostly relaxed on these templates. From the results depicted in Fig. 6 (b) and (d) concludes that the dependence of InAs island lateral size with the local lattice parameter for low dose templates exhibit larger values along a broader range of local lattice parameter conditions. This is a characteristic of partially coherent growth of InAs islands, indicating that, in spite of the presence of defects the crystalline registry with the substrate remains for part of the overall InAs deposited volume. Due to the smaller lateral size of the InAs crystallites in the high dose sample, also confirmed by the GID results of Fig. 6 (b) and (d), InAs islands grow with larger misorientation and reduced coherence in such templates. The InAs peak intensity reduction in the high dose samples diffraction observed in Fig. 6 (a) and (c) is then consequence of the reduction in the volume and quantity of oriented islands grown on top of these substrates, confirming the TEM analysis.

Raman measurements of islands grown on LDT3 and HDT3 conditions were performed at room temperature and backscattering geometry using a triple monochromator Raman spectrometer (model DILOR XY) to study their optical properties. The excitation laser wavelength was 514.5 nm. To avoid sample heating the laser power density was fixed to $10^5$ W/cm$^2$. The scattering peak around 237 cm$^{-1}$ shown on Fig. 7 is ascribed to the longitudinal-optical InAs mode (bulk LO mode is known to be found at 238.6 cm$^{-1}$ [23]). The small observed shift is a consequence of inhomogeneous chemical composition [24] and the size [25]. This peak was observed only in the low-dose sample, which is an additional indication of better InAs crystalline quality on such growth conditions. Comparing the X-ray diffraction results of Fig. 6(a) and (b) with the corresponding Raman spectra it is possible to conclude that the optical quality of InAs islands grown on templates created by ion beams are inversely proportional to the ion dose.

Two additional scattering peaks are observed around 306 and 339 cm$^{-1}$, close to bulk InP values (TO: 304cm$^{-1}$, LO: 345 cm$^{-1}$ [26]). These peaks are assigned to InP transverse-optical (TO) and longitudinal-optical (LO) lattice vibration modes, respectively. The ratio between the InP-TO and InP-LO peak intensities and their widths are related the InP crystal disorder caused by the ion beam interaction with the material [27]. The higher intensity and larger width of the TO peak on the high dose sample is therefore an indication of larger structural disorder of the resulting patterns under such FIB conditions. The Raman measurements point out to a scenario in which the surface damage on the InP caused by the ion beam cannot be reduced with pre-cleaning treatments or during the MOVPE growth. These damages on the InP crystalline structure directly degrade the crystalline quality of the InAs islands grown on high-dose templates and are an inherent drawback of the FIB-patterning process.

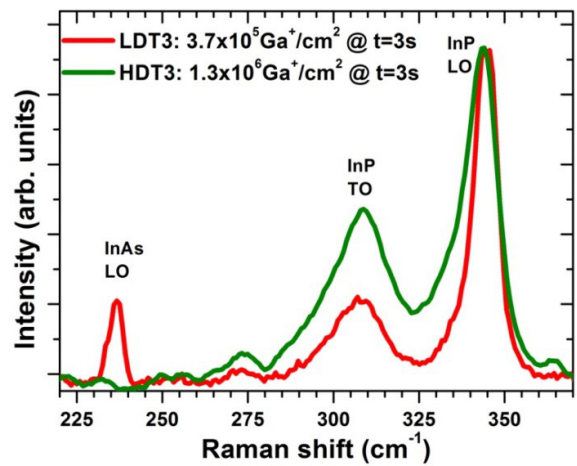

Fig. 7: Raman spectroscopy of LDT3 and HDT3 samples. Only the LDT3 sample exhibits a InAs-LO peak. The stronger and broader TO and LO peaks observed on the HDT3 template are a consequence of the InP crystal damage produced by the ion beam under high-dose conditions.

## 4. DISCUSSION

The growth mechanisms of InAs islands on flat surfaces have been extensively studied along the past decades. For topographically modified surfaces or host crystals with defects there is a relative lack of information to build a common scenario. This is the case for our InP patterned surfaces created by ion beam bombardment in this work. Besides the surface characteristics of the substrate, the temperature and the material flow also affect the growth mechanisms. The non-uniform filling of InAs islands on the cavities (Fig. 2) and the absence of islands in the areas between the holes suggest that the flow of adatoms is the main source of material. However, the flow of precursors that directly impinge cavities coming from the vapor phase is not sufficient for the growth volume observed. A simple analysis of the overall cavity area supports this assertion. The collection area which surrounds the region between holes is 55, 38 and 29 times larger than the cavity cross-sectional area on the surface of InP for the low, medium and high doses, respectively. Hence, the material quantity that is impinging on flat regions is much larger than the amount that directly hits the cross-sectional area of cavities. It is reasonable then to state that the main source of material for the InAs island growth is the net flow of surface adatoms toward to the cavities, pointing out to a strong material transport through the flat regions. This scenario is compatible with the mobility of In atoms at the growth temperatures used here and in agreement with experimental observations for (001) surfaces [28-30].

The islands morphology dependence on the ion dose is then a result of the adatoms flow and the shape (aspect ratio) of cavities. A simplified model of the chemical potential on modified surfaces ($\Delta\mu$) has to be considered in order to describe the adatoms flow [31]. In our case $\Delta\mu$ is given by $\Delta\mu = \mu_S - \mu_S^0 = \Omega\kappa(x) + \Omega E_S(x)$, which is the difference between the chemical potential of the curved surface ($\mu_S$) and flat surface ($\mu_S^0$). $\Omega$ is the atomic volume, $\kappa(x)$ is the surface curvature and $E_S$ is the strain inside the cavity [31,32]. When the surface topography is modified by FIB, the chemical potential is reduced at the cavity edges [32] (see Fig. 7 (a)). Such gradient in $\Delta\mu$ along the surface induces a net rate of adatoms flow towards the cavities [33,34]. At moderate temperatures the adatoms have enough energy to enter cavities but are not able to escape from them, nucleating therein. When the growth occurs at higher temperatures adatoms acquire enough energy to move away from cavities and can eventually nucleate in the regions between the holes [34]. In this latter case, the number of islands formed in the vicinity of the holes would be greater than inside them. The first scenario discussed above was observed in the investigated samples, evidencing that the thermal energy was only enough to induce nucleation on FIB-modified sites. The resulting adatoms flow towards the cavity is larger than the reversal flow (atoms leaving cavities) and most of the precursor atoms nucleate inside, forming the InAs islands.

The observed tendency of adatoms net flow toward the cavities can be understood by a surface chemical potential analysis as well. The equation for $\Delta\mu$ takes into account the surface geometry, which can be considered by assuming that the ion beam has a Gaussian shape. This assumption is corroborated by the Gaussian profile observed on the AFM results of Fig. 1. It is known that the cavity width depends on the ion beam current while its depth depends on the beam total bombardment time and on the sputtering rate of the substrate material [35,36]. Therefore, $\Delta\mu$ must show a direct dependency with ion dose, which in turn determines the surface curvature. These considerations point out to a scenario in which the adatoms net flow towards the cavities is larger for deeper cavities (high-dose templates). Although the surface chemical potential model supports the basic idea of material mobility towards the cavities, it is unable to describe the island morphology and density dependence on the ion dose. According to this model, the islands are expected to be bigger for larger adatoms flow into the cavities. However, such configuration was not observed in the SEM images and GID results on high-dose templates, where smaller islands with increased density were found (Fig. 2). This result can only be explained based on the larger density of structural defects observed by Raman scatering on the high-dose templates. Under high-dose conditions a larger density of small islands tends to grow with a stronger misorientation with respect to the substrate and grow with incoherently. In these templates a nucleation regime is favored with respect a coarsening mechanism.

On low-dose substrates a reduced amount of InP defects is observed and cavities present reduced surface area. In such conditions the average distance between the critical nuclei [37], formed in the early stages of growth, is smaller than in the high dose cavities. As a result coarsening is favored since the distance between the critical nuclei is shorter than the adatoms diffusion length [38]. Some islands grown on the low-dose templates exhibit crystalline defects as seen in the TEM images and most of them grow in a partially-coherent state according to the statistically significant results from X-ray grazing-incidence diffraction. The conditions depicted in the paragraphs above explain both the increase in the diffraction signal of InAs islands and the larger lateral size observed in GID for islands grown on low-dose templates and agrees with the shape and crystallographic orientations of the islands observed in TEM images.

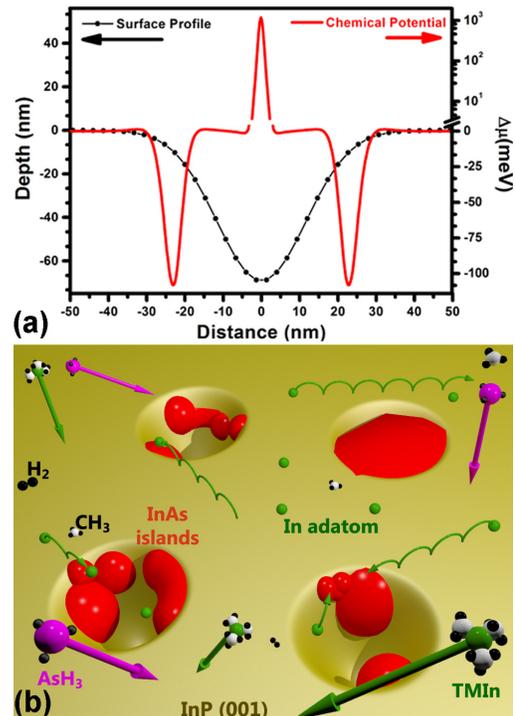

Fig. 8: Sketch of the island growth process on templates created by focused ion beam. In (a) the dotted curve shows a model of the cavity profile and the red solid curve depicts the surface chemical potential dependence with the surface curvature. In such conditions the surface chemical potential is reduced at the edges of the cavities. The existence of such condition and the adatom mobility at the growth temperatures used result in a net flow of adatoms towards the cavity. In (b) a schematic diagram of the adatom diffusion process into cavities and the island growth is presented. The adatoms that impinge flat areas are able to diffuse and attach to the cavities depending on growth conditions (temperature, precursor pressure). Once adatoms enter the cavity their nucleation and the consequent island shape depends on the defect density created due to gallium ion bombardment.

## 5. CONCLUSION

In this work the growth of InAs islands on InP(100) surfaces patterned by FIB was investigated. Regular arrays of cavities were produced under three different ion doses and InAs islands deposited with two deposition times (3 s and 4 s) exposure to the In precursor. Scanning electron microscopy, transmission electron microscopy and grazing incidence X-ray diffraction were employed to investigate the structure of the grown islands. The formation of InAs islands and small clusters inside most of the cavities were observed, while no islands were detected in flat surfaces. Large InAs crystals filling completely some of the cavities were found in the low-doses (LD) samples. GID results reveal that these islands are partially coherent. High-doses (HD) templates present mainly incoherent (relaxed) InAs nanocrystals. A simple model was proposed to correlate the chemical potential of the modified surface and the flow of adatoms. The differences between the LD and HD samples are ascribed to the relation between the average nuclei-nuclei distance due to the cavity size, the density of defects on InP due to ion bombardment and the diffusion length of adatoms. On LD samples the reduced number of surface defects and the samller cavity size favors island coarsening while nucleation processes dominate the formation of InAs islands on HD templates.

## 6. ACKNOWLEDGEMENTS

This work was supported by the INCT-DISSE, the Brazilian agencies CNPq, CAPES, FAPEMIG and FAPERJ. We would like to thank the Microscopy Center of the Universidade Federal de Minas Gerais as well the Brazilian Synchrotron Light Laboratory (LNLS) under proposal XRD2-11795 for their technical and financial support of this work. We would like to thank also the Nano Spectroscopy Laboratory (http://www.labns.com.br) as well as the Prof. Dr. Luiz Gustavo Cançado for helping with Raman measurements. We are also grateful to M. Sc. Wesller Schmidt for helping with the GID experiments.